\documentclass{aa}     
\usepackage{graphics}

\def\cm2{\,{\rm cm^{-2}}}

\def\13co{\,{\rm ^{13}CO}}
\def\h2{\,{\rm H_{2}}}
\def\aua{{\rm A\&A} }
\def\auas{{\rm A\&AS} }
\def\aar{{\rm A\&A Rev.} }
\def\apj{{\rm ApJ} }
\def\aj{{\rm AJ} }
\def\apjs{{\rm ApJS} }
\def\apjl{{\rm ApJL} }

\def\mnras{{\rm MNRAS} }
\def\pasj{{\rm PASJ} }
\def\pasp{{\rm PASP} }

\begin{document}

\title{The millimeter-wave continuum spectrum of Centaurus A and its nucleus}
 
   \subtitle{}
 
\author{F.P. Israel \inst{1},
        D. Raban \inst{1},
        R.S. Booth, \inst {2,3}
        and F.T. Rantakyr\"o \inst{4}   }
 
   \offprints{F.P. Israel}
 
  \institute{Sterrewacht Leiden, Leiden University, P.O. Box 9513, 2300 RA 
             Leiden, The Netherlands
   \and      Onsala Space Observatory, Chalmers University of Technology, 
             SE-439 92 Onsala, Sweden
   \and      Hartebeesthoek Radio Astronomy Observatory, PO Box 443, 
             Krugersdorp 1740, South Africa
   \and      European Southern Observatory,  Casilla 19001, Santiago 19, 
             Chile
}
 
\date{Received ????; accepted ????}
 
\abstract{} {We study the radio emission mechanism of the FR-I AGN
  NGC~5128 (Centaurus A)} {We determine the centimeter and
  millimeter continuum spectrum of the whole Centaurus A radio source
  and measure at frequencies between 86 GHz (3.5 mm) and 345 GHz (0.85
  mm) the continuum emission from the active radio galaxy nucleus at
  various times between 1989 and 2005.}  {The integral radio source
  spectrum becomes steeper at frequencies above 5 GHz, where the
  spectral index changes from $\alpha_{low}$ = -0.70 to
  $\alpha_{high}$ = -0.82.  The SW outer lobe has a steeper spectrum
  than the NE middle and outer lobes ($\alpha$ = -1.0 vs -0.6).
  Millimeter emission from the core of Centaurus A is variable, a
  variability that correlates appreciably better with the 20-200 keV
  X-ray variability than with 2 - 10 keV variability.}  {In its
  quiescent state, the core has a spectral index $\alpha$ = -0.3,
  which steepens when the core brightens. The variability appears to
  be mostly associated with the inner nuclear jet components that have
  been detected in VLBI measurements. The densest nuclear components
  are optically thick below 45-80 GHz.  }

\titlerunning{Centaurus A millimeter continuum emission}
\maketitle

\section{Introduction}

\begin{figure*}[]
\unitlength1cm
\begin{minipage}[t]{3.6cm}
\resizebox{5.7cm}{!}{\rotatebox{0}{\includegraphics*{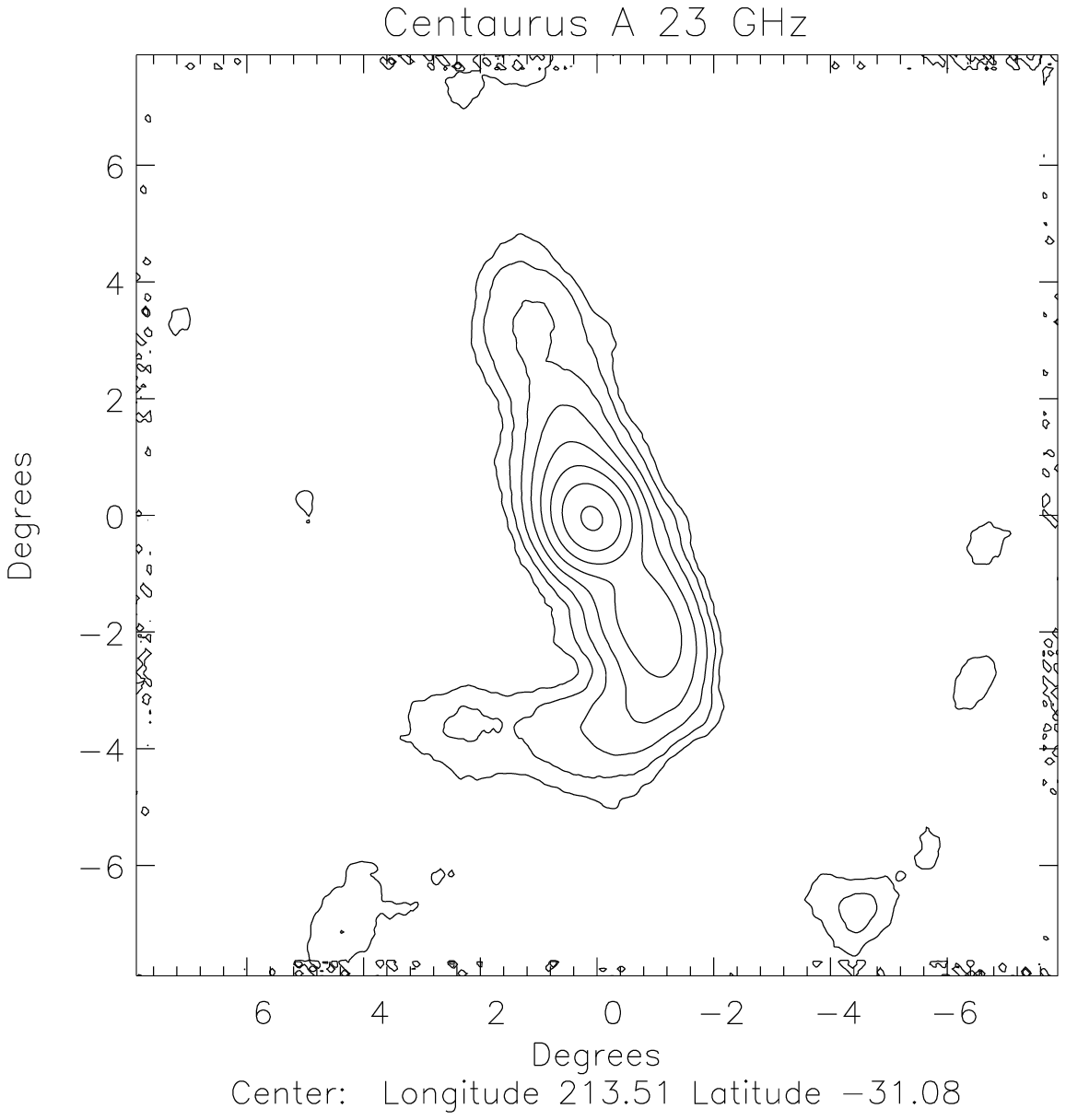}}}
\end{minipage}
\begin{minipage}[t]{3.6cm}
\resizebox{5.7cm}{!}{\rotatebox{0}{\includegraphics*{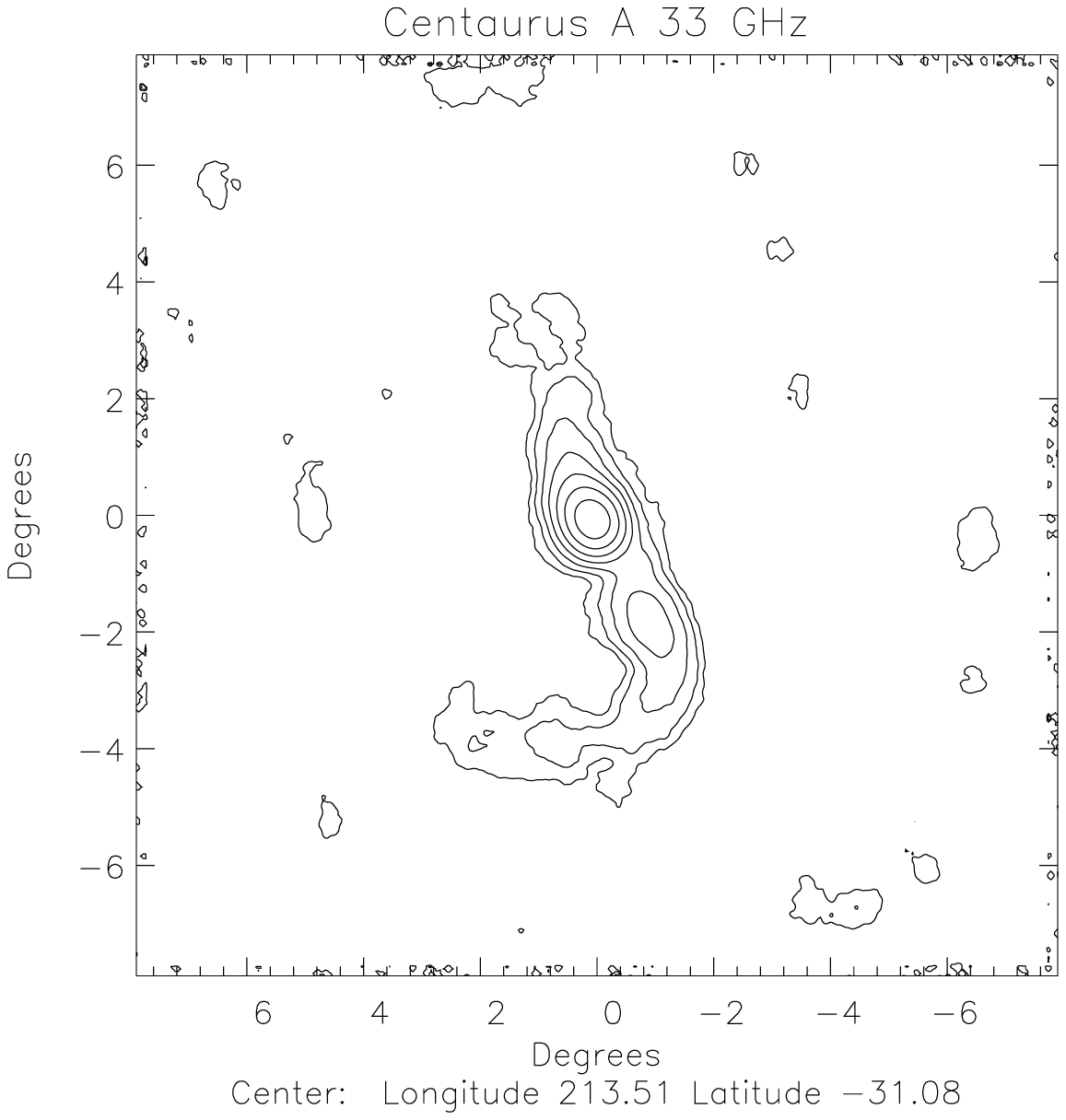}}}
\end{minipage}
\begin{minipage}[t]{3.6cm}
\resizebox{5.7cm}{!}{\rotatebox{0}{\includegraphics*{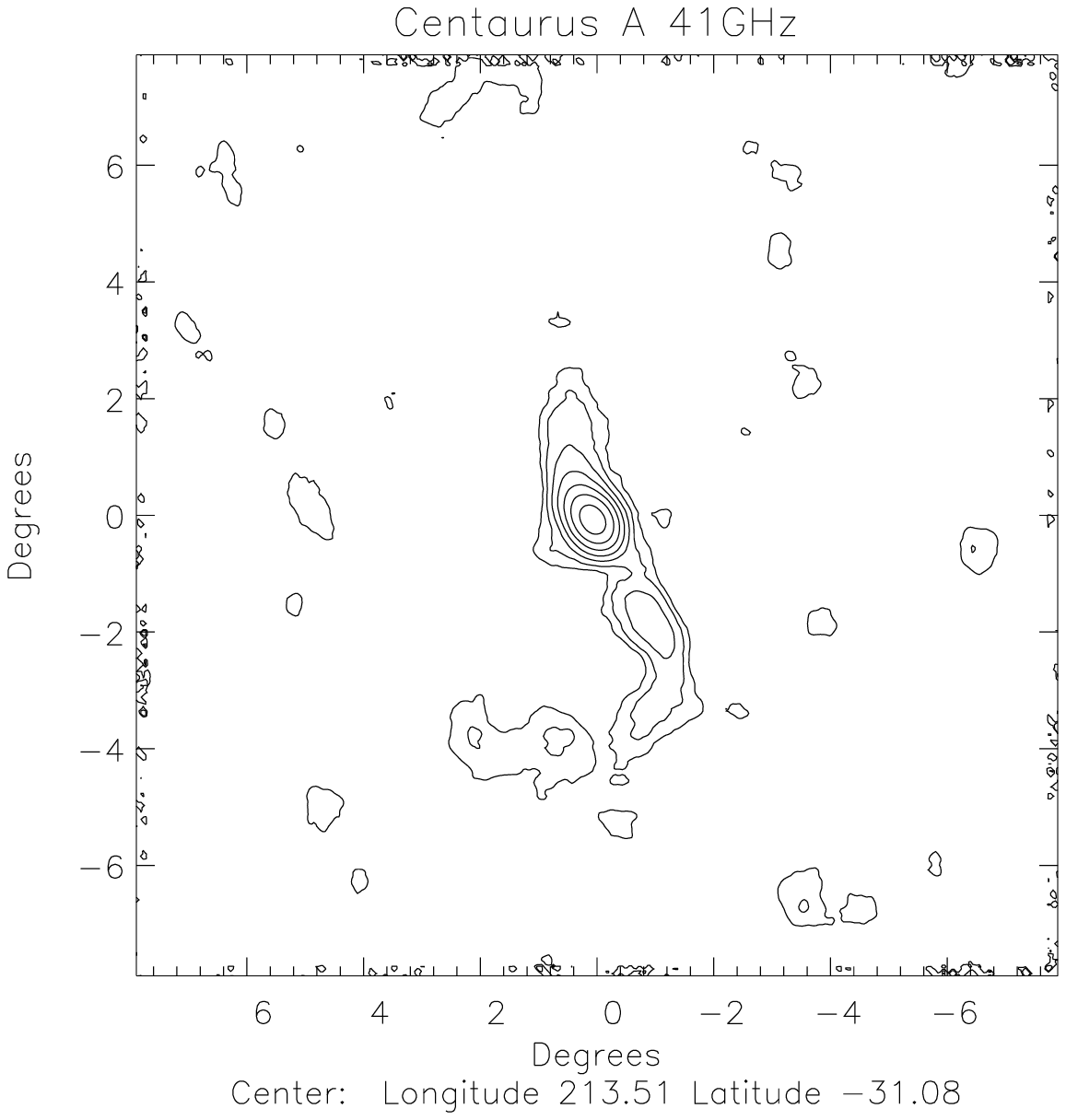}}}
\end{minipage}
\begin{minipage}[t]{3.6cm}
\resizebox{5.7cm}{!}{\rotatebox{0}{\includegraphics*{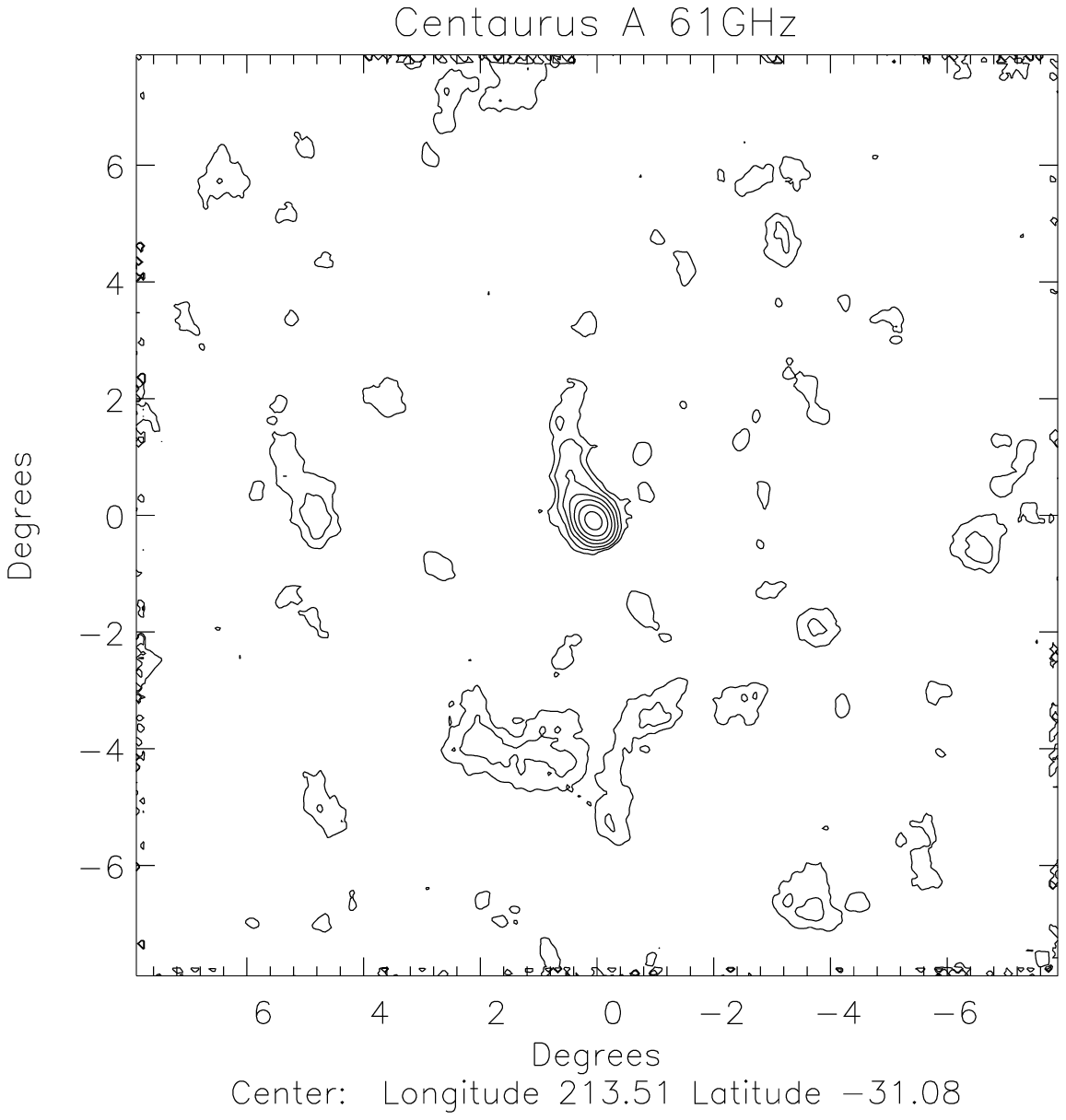}}}
\end{minipage}
\begin{minipage}[t]{3.6cm}
\resizebox{5.7cm}{!}{\rotatebox{0}{\includegraphics*{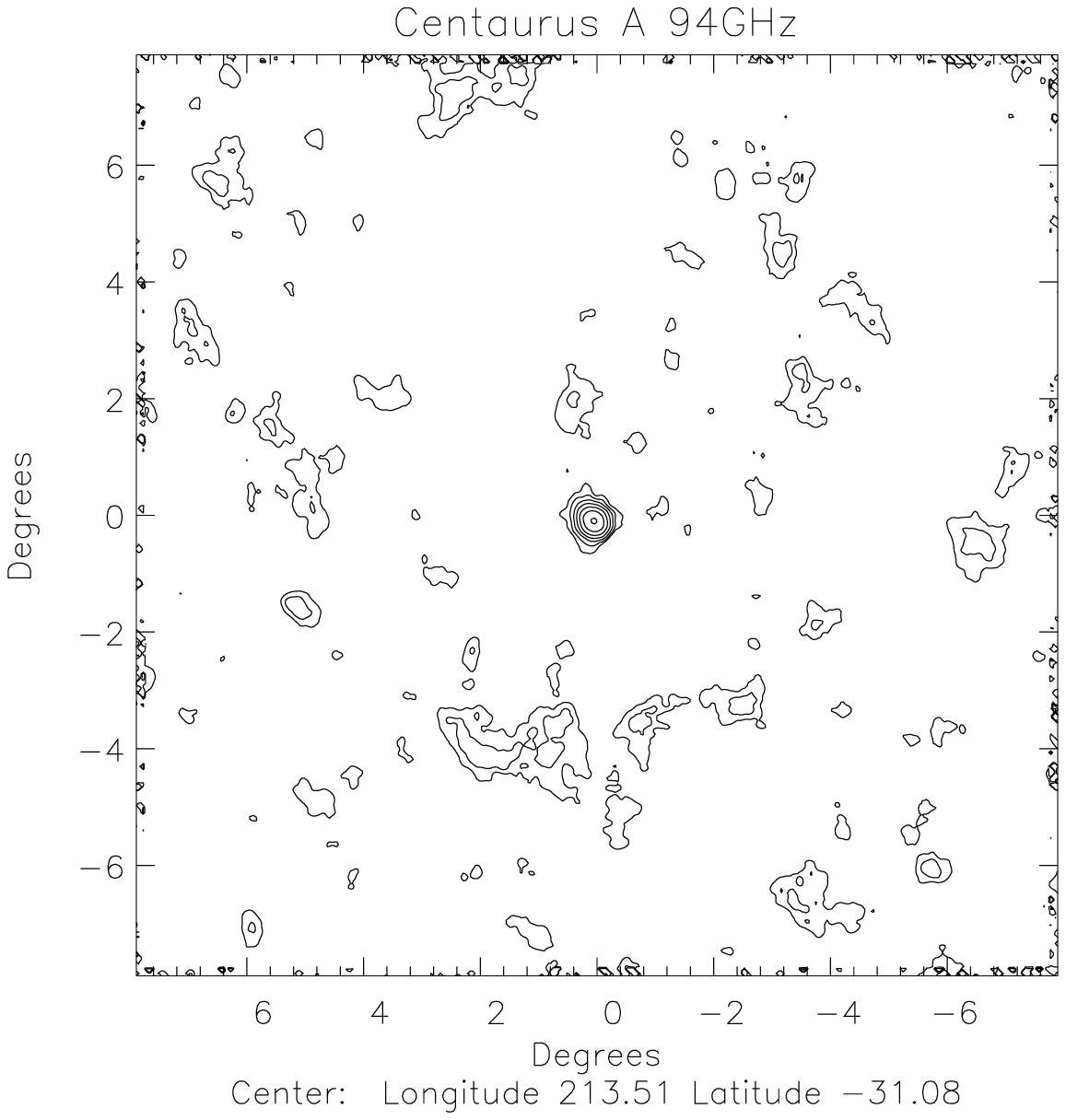}}}
\end{minipage}
\caption[]{Maps of the radio continuum emission of Centaurus A at
  (left to right) 23, 33, 41 GHz, 61 GHz, and 93 GHz.  All images are
  at the nominal WMAP resolution.  Contour levels are drawn at (23
  GHz) 0.1, 0.19, 0.38, 0.74, 1.4, 2.8, 5.5, 10.7 mK, (33 GHz) 0.1,
  0.19, 0.36, 0.68, 1.3, 2.4, 4.6 mK; (41 GHz) 0.1, 0.19, 0.34, 0.64,
  1.2, 2.2, 4.1, 7.6 mK; (61 GHz) 0.1, 0.17, 0.31, 0.53, 0.94, 1.6,
  2.9, 5.0 mK; (93 GHz) 0.1, 0.16, 0.25, 0.40 0.64 1.0, 1.6, 2.6
  mK. Note the persistent diffuse emission features at +1.5$^{\circ}$,
  -4$^{\circ}$ and at +2$^{\circ}$, +8$^{\circ}$, which are probably
  unrelated Galactic foreground emission (see Sect. 2.1.).}
\begin{minipage}[t]{3.6cm}
\resizebox{4.15cm}{!}{\rotatebox{0}{\includegraphics*{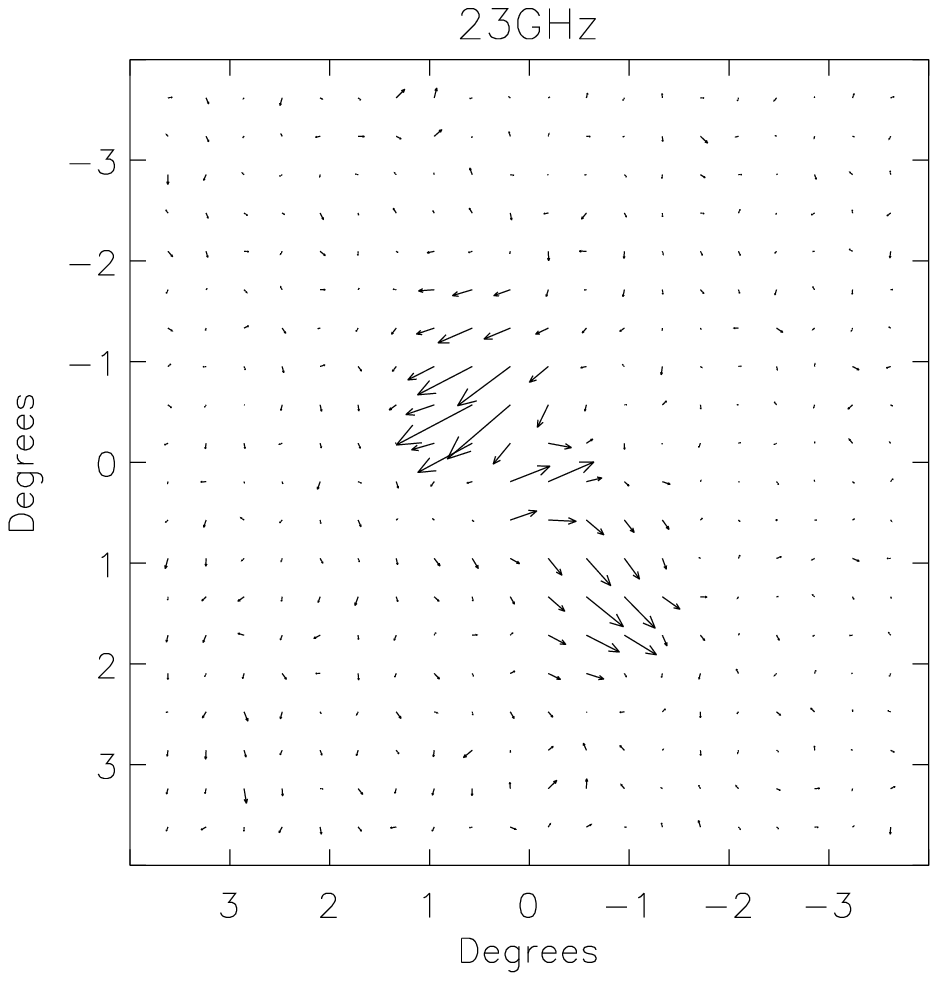}}}
\end{minipage}
\begin{minipage}[t]{3.6cm}
\resizebox{4.15cm}{!}{\rotatebox{0}{\includegraphics*{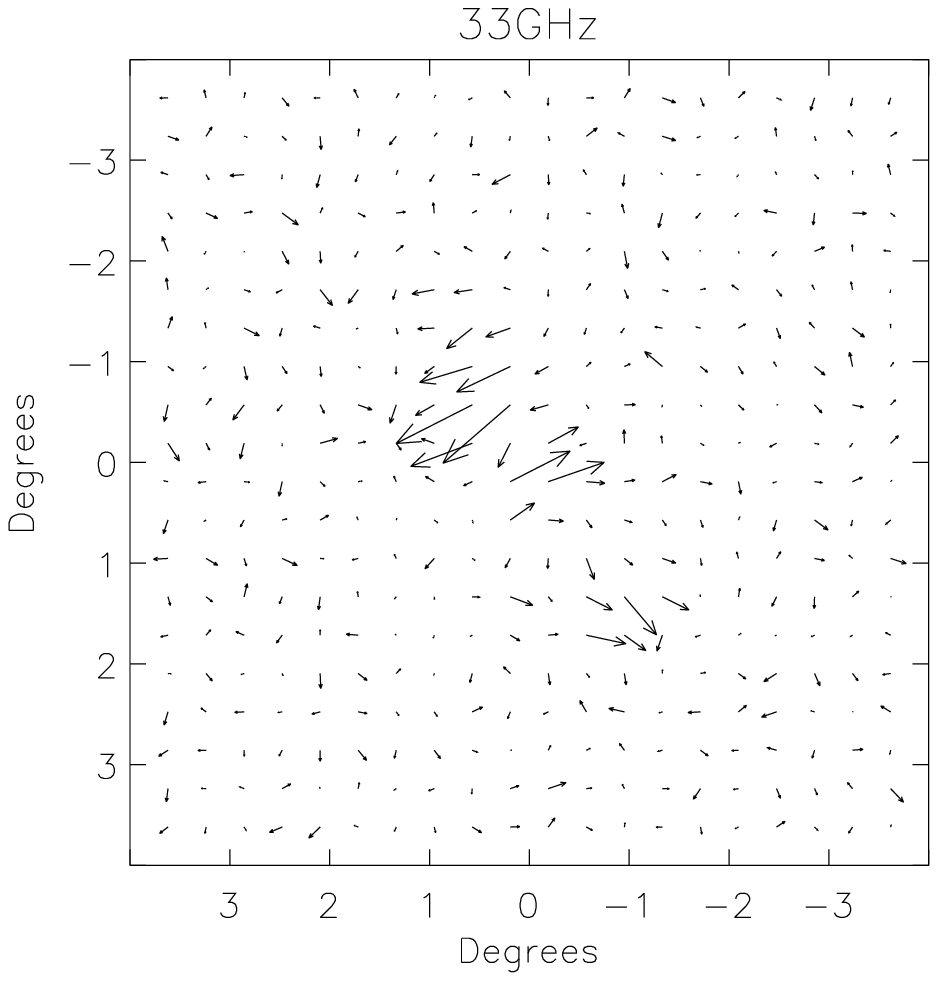}}}
\end{minipage}
\begin{minipage}[t]{3.6cm}
\resizebox{4.15cm}{!}{\rotatebox{0}{\includegraphics*{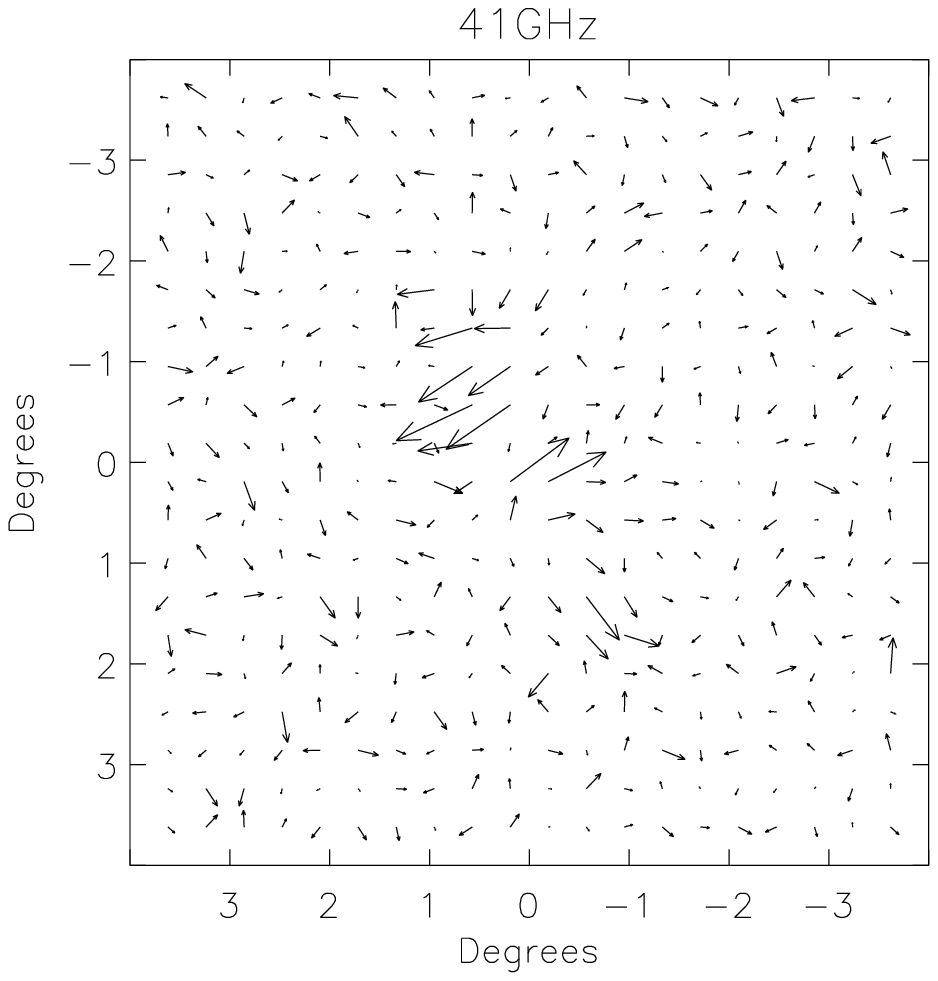}}}
\end{minipage}
\begin{minipage}[t]{3.6cm}
\resizebox{4.15cm}{!}{\rotatebox{0}{\includegraphics*{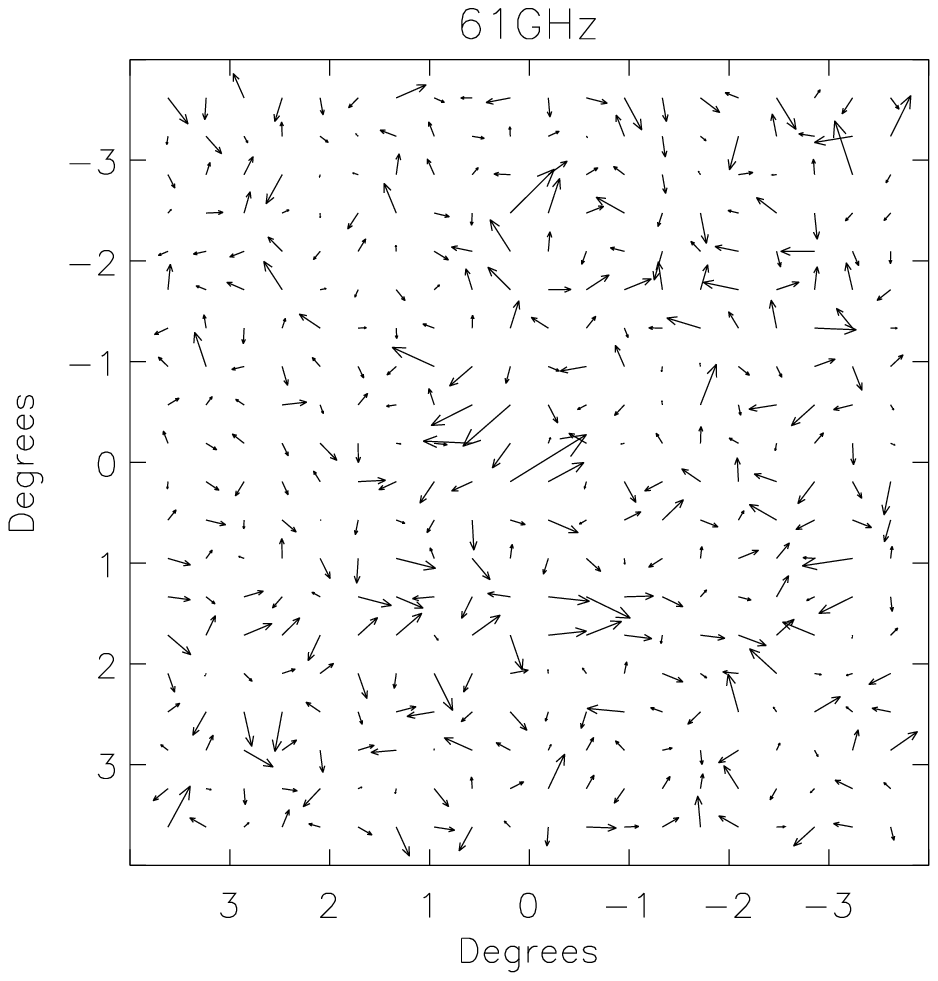}}}
\end{minipage}
\begin{minipage}[t]{3.6cm}
\resizebox{3.9cm}{!}{\rotatebox{0}{\includegraphics*{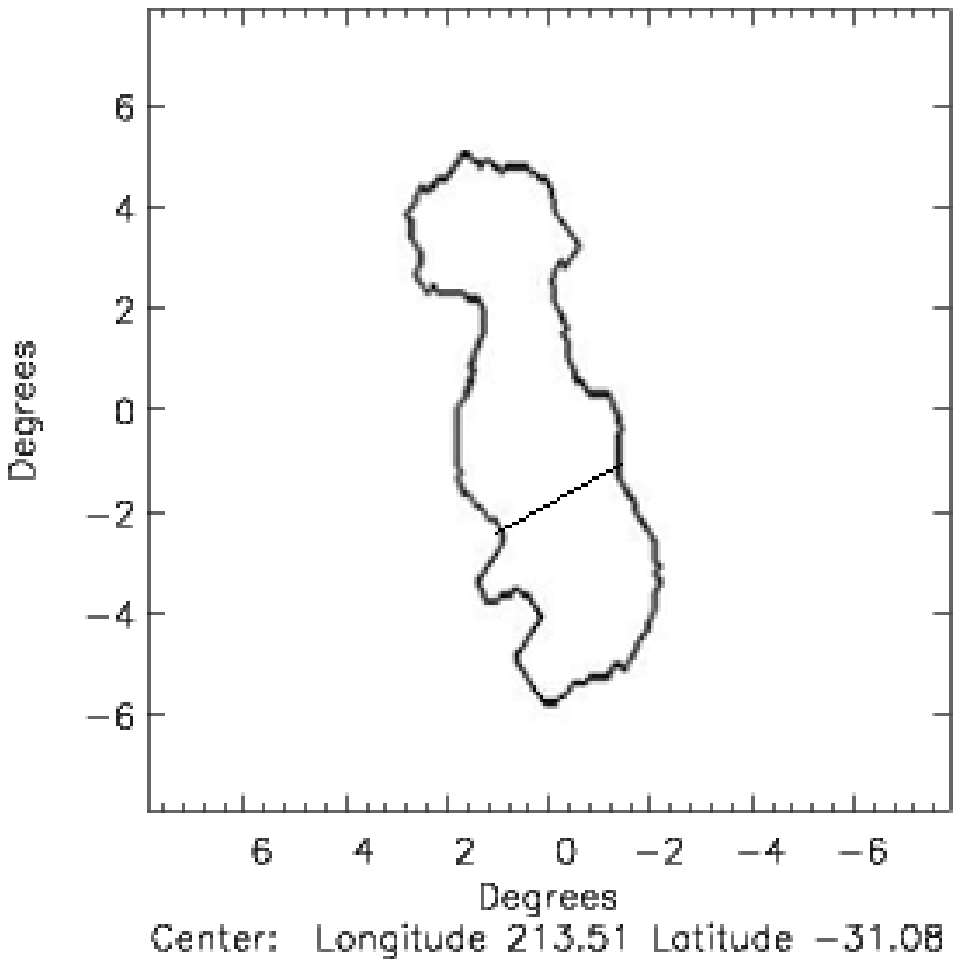}}}
\end{minipage}
\caption[]{Polarized vectors of the Centaurus A emission at (left to
  right) 23, 33, 41 GHz, and 61 GHz; the length of the vectors is
  proportional to the polarized intensity, but scaling is
  arbitrary. Far right: masks used to determine flux densities in
  Table\,\ref{WMAPdata}.  }
\label{WMAPs}
\end{figure*}

\begin{table*}
\caption[]{WMAP observations of Centaurus A}
\begin{center}
\begin{tabular}{ccccccc}
\hline
\noalign{\smallskip}
Frequency & \multicolumn{3}{c}{Flux Densities} & \multicolumn{3}{c}{Polarized Flux Densities} \\
     & Total  & NE Lobes+Core & SW Lobes & Total  & NE Lobes+Core & SW Lobes \\
(GHz)     & \multicolumn{6}{c}{Jy}\\
\noalign{\smallskip}     
\hline
\noalign{\smallskip}
 23  &112$\pm$13 & 76$\pm$9  & 29$\pm$4  & 6.8$\pm$0.5 & 5.0$\pm$0.3 & 1.8$\pm$0.2 \\
 33  & 82$\pm$12 & 62$\pm$8  & 20$\pm$4  & 4.6$\pm$0.6 & 3.6$\pm$0.3 & 0.8$\pm$0.2 \\
 41  & 72$\pm$9  & 56$\pm$6  & 16$\pm$3  & 3.8$\pm$0.5 & 3.4$\pm$0.5 & 0.4$\pm$0.2 \\
 61  & 56$\pm$15 & 48$\pm$12 &  8$\pm$3  & 4.0$\pm$1   & 4.0$\pm$1   &     ---     \\
 93  & 41$\pm$18 & 47$\pm$18 & -6$\pm$9  & 3.5$\pm$1   & 3.5$\pm$1   &     ---     \\
\noalign{\smallskip}
\hline
\end{tabular}
\end{center}
\label{WMAPdata}
\end{table*}
 
Centaurus A (NGC~5128) is the nearest (D = 3.4 Mpc) radio galaxy with
an active nucleus.  Because of this, it has been extensively studied
at various wavelengths over a very wide range in the electromagnetic
spectrum; much of the observational database was reviewed by
Israel (1998).  The radio source spans more than eight degrees on the
sky and shows structure on all scales down to milliarcseconds where
VLBI techniques (e.g.  Horiuchi et al. 2006) reveal an extremely
compact core and bright nuclear jets.  Notwithstanding the prominence
of Centaurus A, relatively few measurements of the total radio
emission of the source exist.  The first comprehensive study of its
decimetric emission and polarization by Cooper et al (1965) has, in
more recent times, been supplemented only by the measurements
summarized by Alvarez et al. In particular, no measurements of the
Centaurus A total flux density at frequencies above 5 GHz are found in
the astronomical literature.  In this paper, we extend that range by
providing new determinations between 20 GHz and 100 GHz derived from
the WMAP all-sky surveys.

On much smaller scales, the flux densities of the compact Centaurus A
core are also of interest. The central region of Centaurus A has been
imaged at very high resolution with (space) VLBI techniques (Meier et
al. 1989; Tingay et al. 1998, Fujisawa et al, 2000, Tingay, Preston
$\&$ Jauncey 2001, Horiuchi et al. 2006) up to frequencies of 22 GHz,
where both the selfabsorbed synchrotron emission from the core and the
free-free absorbing foreground start to become optically thin (Tingay
$\&$ Murphy 2001).  Although Kellerman, Zensus $\&$ Cohen (1997) used
the VLBA to observe Centaurus A at the significantly higher frequency
43 GHz, they did not actually measure nuclear flux densities.
Hawarden et al. (1993) have provided the most recent systematic
attempt to derive the core spectrum from single-dish measurements, and
they conclude that it is essentially flat at millimeter wavelengths, a
conclusion reiterated by Leeuw et al (2002).

Variability of the emission at frequencies of 22 and 43 GHz was
explicitly noticed by Botti $\&$ Abraham (1993), de Mello $\&$ Abraham
(1990), and Abraham et al. (2007). However, these results, as well as
those by Tateyama $\&$ Strauss (1992), generally suffered from low
spatial resolutions of 2--4 arcmin that do not allow separate
identification of emission from the nucleus, the jets, and their
surroundings.  In this paper, we provide new flux density measurements
of the compact core in the 90--345 GHz (sub)millimeter range, at
resolutions between 0.8 and 0.2 arcmin.  As these measurements were
collected over a time period spanning more than a decade, we are also
in a position to study the variable behavior of this emission.

\section{Observations and reduction}

\subsection{WMAP data}

The WMAP mission and its data products have been described in detail
by Bennett et al (2003a, b, c).  In our analysis, we used the data
products from the official WMAP 5-year release, specifically the fully
reduced and calibrated Stokes I, Q, and U maps of the entire sky, in
nested HEALPIX format (Hinshaw et al. 2008).  The maps were observed
at frequencies (assuming synchrotron spectra) $\nu$ = 22.5, 32.7,
40.6, 60.7, and 93.1 GHz with resolutions of 53, 40, 31, 21, and 13
arcmin respectively.  As a first step, the HEALPIX data maps were
converted with the use of standard software to flat maps in Zenithal
Equal Area projection with pixel solid angles of
$p(o)\,=\,1.90644\,\times\,10^{-6}$ and intensities in mK.  We first
attempted to determine integrated flux densities by fitting in each
map two-dimensional Gaussians to the emission corresponding to
Centaurus A. However, with increasing frequency, both the beamsize and
the total flux density decrease, resulting in a steep decline in flux
density per beam.  Our fitting results, not surprisingly, suggests
that we were systematically missing low-surface-brightness extended
emission.  For this reason, we also integrated over the full extent of
the Centaurus A radio source as shown in the maps by Cooper et
al. (1965) - the mask used is depicted in Fig.\,\ref{WMAPs}, bottom
right. We derived the flux densities given in Table\,\ref{WMAPdata}
from the summed values with the conversion factor Jy/mK = 30.7
$\times$ $p(o)$ $\times$ $\nu^{2}$.  This yielded the flux densities
given in Table\,\ref{WMAPdata}, which correspond to the maps shown in
Fig.\,\ref{WMAPs}.  We verified the accuracy of this procedure by also
applying it to the strongest WMAP point sources J0319+4131,
J0322-3711, J0423-0120, J1229-0203, J1230+1223, and J1256-0547 (Wright
et al. 2008). For polarization measurements, the Q and U maps were
reduced using the same conversion factors as the I maps; the polarized
intensity was computed as $P=\sqrt{Q^2+U^2}$.  Only at 23 GHz are
signal-to-noise ratios high enough to yield practically identical
results for the gaussian-fitting and integration methods.  We consider
the integrated flux densities at 23, 33, and 41 GHz to be quite
reliable, whereas at 61 GHz and 93 GHz much of the integration was
over areas with a surface brightness very close to the noise,
especially in the southern lobe.

In the total-intensity (I) maps, the spatial resolution blends all
components into the dominant emission of the NE and SW radio lobes,
respectively.  Even at the highest frequency, the compact core cannot
be separated from the intense NE lobe. This is different in the
polarized intensity (P) maps. In the 23, 33, and 41 GHz P-maps, the
compact core can be seen separately. At the higher frequencies, the
signal-to-noise ratio of the P-maps is too low to yield useful
results.  We note that the {\it central} 22 GHz flux densities of Cen
A typically range between 16 and 29 Jy in a $4'$ beam (Fogarty $\&$
Schuch 1975; Botti $\&$ Abraham 1993), i.e.  roughly 15--25$\%$ of the
23 GHz WMAP flux densities. At the same frequency, the VLBI
milli-arcsec core (Tingay et al. 2001) contributes only $2\%$ to the
measured WMAP emission.

Figure\,\ref{WMAPs} shows the presence of an extended diffuse emission
feature to the east of the SW lobe.  This feature can be seen at all
frequencies and is undoubtedly real.  However, we do not consider it
to be related to Centaurus A. (1) Large-scale maps show significant
foreground emission from the Milky Way reaching up to the Galactic
latitude of Centaurus A (see Bennett et al. 2003b). (2) Inclusion of
this feature in the flux densities of the SW lobe causes its emission
to first decrease with frequency but then to make a surprising jump at
61 GHz, and at 93 GHz the weak SW lobe would have become as strong as
the NE lobe + core emission. (3) The lobes of Centaurus A are clearly
polarized, whereas the diffuse feature does not show polarization at
any frequency (Fig.\,\ref{WMAPs}, bottom row), which also argues
against the feature being an extension of the SW lobe. Finally, (4)
there is a similar but more obviously unrelated feature at 
+2$^{\circ}$, 8$^{\circ}$.

\subsection{SEST data}

The SEST was a 15 m radio telescope located at La Silla, Chile.
\footnote{The Swedish-ESO Submillimeter Telescope was operated jointly
  by the European Southern Observatory (ESO) and the Swedish Science
  Research Council (NFR).} The main purpose of our observing program
with the SEST was to obtain high-resolution molecular-line absorption
spectra against the continuum of the Centaurus A core, with integration
times ranging from 30 minutes to 4 hours per molecular transition per
session, obtained by co-adding individual samples usually of 5 minutes
each.

\begin{table*}
\caption[]{SEST observations of the Centaurus A nucleus}
\begin{center}
\begin{tabular}{lcccclcccc}
\hline
\noalign{\smallskip}
Day      & Date   & Frequency & Brightness & Flux Density & Day & Date   & Frequency & Brightness & Flux Density \\
         &        &$\nu$ (GHz)&T$_{\rm mb}$ (mK)&S$_{\nu}$ (Jy)&&        & $\nu$ (GHz) &T$_{\rm mb}$ (mK)& S$_{\nu}$ (Jy)         \\
\noalign{\smallskip}     
\hline
\noalign{\smallskip}
1989.29 & Apr 15 &  87 & 373$\pm$45 &  7.2$\pm$0.9 & 1998.61 & Aug 15 &  90 & 286$\pm$18 &  7.1$\pm$0.7 \\ 
1989.50 & Jul 3  &  89 & 391$\pm$15 &  7.3$\pm$0.7 &         &        & 230 & 138$\pm$10 &  5.7$\pm$0.6 \\
        &        &  98 & 393$\pm$7  &  7.5$\pm$0.8 & 1998.99 & Dec 30 &  90 & 321$\pm$13 &  8.0$\pm$0.8 \\
        &        & 110 & 313$\pm$20 &  5.9$\pm$0.6 &         &        & 150 & 226$\pm$16 &  6.8$\pm$0.7 \\
        &        & 113 & 353$\pm$9  &  6.6$\pm$0.7 & 1999.10 & Feb 04 &  90 & 277$\pm$16 &  6.9$\pm$0.7 \\
        &        & 220 & 223$\pm$24 &  5.0$\pm$0.7 &         &        & 150 & 184$\pm$20 &  5.5$\pm$0.6 \\
        &        & 227 & 208$\pm$44 &  4.3$\pm$0.8 & 1999.12 & Feb 11 &  90 & 304$\pm$16 &  7.6$\pm$0.8 \\
1991.03 & Nov 01 &  88 & 371$\pm$20 &  7.1$\pm$0.8 &         &        & 150 & 283$\pm$37 &  8.5$\pm$1.2 \\
        &        &  89 & 369$\pm$21 &  7.0$\pm$0.8 & 1999.13 & Feb 18 &  90 & 333$\pm$24 &  8.3$\pm$0.6 \\
        &        &  91 & 361$\pm$20 &  6.9$\pm$0.8 &         &        & 150 & 282$\pm$33 &  8.5$\pm$1.0 \\
        &        & 114 & 361$\pm$25 &  6.8$\pm$0.9 & 1999.38 & May 17 &  86 & 372$\pm$45 &  7.0$\pm$0.9 \\
1992.53 & Jul 13 &  89 & 516$\pm$11 &  9.7$\pm$1.0 &         &        &  90 & 373$\pm$45 &  7.1$\pm$0.9 \\
        &        &  91 & 516$\pm$7  &  9.7$\pm$1.0 &         &        &  98 & 362$\pm$45 &  6.9$\pm$0.7 \\
        &        &  99 & 488$\pm$31 &  9.2$\pm$1.2 &         &        & 110 & 377$\pm$45 &  7.2$\pm$0.7 \\ 
        &        & 218 & 263$\pm$11 &  5.5$\pm$0.6 &         &        & 169 & 168$\pm$45 &  5.5$\pm$0.6 \\
        &        & 220 & 326$\pm$60 &  6.8$\pm$1.3 &         &        & 226 & 276$\pm$45 &  5.5$\pm$0.6 \\
        &        & 226 & 289$\pm$35 &  5.8$\pm$0.8 &         &        & 230 & 250$\pm$45 &  5.1$\pm$0.5 \\
        &        & 268 & 210$\pm$13 &  4.8$\pm$0.7 &         &        & 267 & 190$\pm$45 &  4.4$\pm$0.8 \\
1993.55 & Jul 20 & 218 & 384$\pm$40 &  8.0$\pm$0.8 & 1999.49 & Jun 26 &  90 & 394$\pm$14 &  9.8$\pm$1.0 \\
        &        & 226 & 354$\pm$40 &  7.3$\pm$0.7 &         &        & 150 & 297$\pm$16 &  8.9$\pm$0.9 \\
        &        & 268 & 329$\pm$37 &  7.6$\pm$0.8 & 1999.61 & Aug 12 &  90 & 299$\pm$18 &  7.5$\pm$0.8 \\
        &        & 346 & 292$\pm$30 &  7.6$\pm$0.8 &         &        & 115 & 285$\pm$40 &  7.7$\pm$1.1 \\
        &        & 357 & 238$\pm$35 &  5.6$\pm$0.6 &         &        & 150 & 299$\pm$18 &  9.0$\pm$0.9 \\
1996.67 & Aug 31 & 110 & 353$\pm$17 &  6.7$\pm$0.7 & 1999.98 & Dec 25 &  90 & 292$\pm$25 &  7.3$\pm$0.7 \\
        &        & 115 & 333$\pm$31 &  6.3$\pm$0.6 &         &        & 150 & 250$\pm$26 &  7.5$\pm$0.8 \\
        &        & 145 & 275$\pm$14 &  5.5$\pm$0.6 & 2000.37 & May 15 &  85 & 408$\pm$21 &  7.8$\pm$0.8 \\
1996.97 & Dec 19 &  90 & 410$\pm$20 & 10.3$\pm$1.0 &         &        &  89 & 397$\pm$19 &  7.4$\pm$0.8 \\
        &        & 229 & 200$\pm$20 &  8.2$\pm$0.8 &         &        &  98 & 374$\pm$24 &  7.2$\pm$0.7 \\
1996.99 & Dec 29 &  90 & 380$\pm$20 &  9.5$\pm$1.0 &         &        & 111 & 374$\pm$22 &  7.1$\pm$0.7 \\
        &        & 229 & 160$\pm$20 &  6.6$\pm$1.6 &         &        & 219 & 233$\pm$20 &  4.8$\pm$0.5 \\
1997.10 & Feb 06 &  90 & 350$\pm$10 &  8.8$\pm$0.9 &         &        & 262 & 205$\pm$68 &  4.7$\pm$0.5 \\
        &        & 150 & 250$\pm$20 &  7.5$\pm$0.8 &         &        & 267 & 227$\pm$30 &  5.2$\pm$0.6 \\
1997.15 & Feb 21 &  90 & 370$\pm$10 &  9.3$\pm$0.9 & 2002.07 & Jan 25 &  87 & 521$\pm$30 &  9.8$\pm$1.0 \\
        &        & 150 & 290$\pm$10 &  8.7$\pm$0.9 &         &        &  91 & 495$\pm$25 &  9.4$\pm$0.9 \\
1997.30 & Apr 19 &  90 & 360$\pm$30 &  9.0$\pm$0.9 &         &        & 109 & 406$\pm$50 &  8.6$\pm$1.0 \\
1997.68 & Sep 05 &  87 & 440$\pm$20 &  8.3$\pm$0.8 &         &        & 267 & 243$\pm$70 &  5.5$\pm$1.6 \\
        &        &  89 & 420$\pm$50 &  7.9$\pm$0.9 & 2002.36 & May 10 &  84 & 482$\pm$11 &  8.9$\pm$0.9 \\
        &        & 113 & 353$\pm$55 &  6.6$\pm$0.8 &         &        &  85 & 477$\pm$5  &  9.0$\pm$0.9 \\
        &        & 145 & 318$\pm$30 &  6.3$\pm$0.6 &         &        &  87 & 475$\pm$27 &  8.9$\pm$0.9 \\
        &        & 150 & 393$\pm$30 &  6.0$\pm$0.6 &         &        &  88 & 459$\pm$23 &  8.6$\pm$0.9 \\
        &        & 220 & 265$\pm$50 &  5.4$\pm$1.0 &         &        &  89 & 476$\pm$15 &  8.9$\pm$0.9 \\
1998.49 & Jun 28 &  90 & 260$\pm$20 &  6.5$\pm$0.7 &         &        &  93 & 511$\pm$14 &  9.5$\pm$1.0 \\
        &        & 150 & 300$\pm$20 &  9.0$\pm$0.9 &         &        &  97 & 461$\pm$15 &  8.8$\pm$0.9 \\
1998.50 & Jul 03 &  89 & 424$\pm$20 &  8.0$\pm$0.8 &         &        & 109 & 438$\pm$25 &  8,3$\pm$0.8 \\
        &        &  98 & 370$\pm$20 &  7.0$\pm$0.7 &         &        & 147 & 375$\pm$17 &  7.4$\pm$0.7 \\
        &        & 218 & 335$\pm$20 &  6.9$\pm$0.7 &         &        & 218 & 213$\pm$17 &  4.4$\pm$0.5 \\
        &        & 227 & 212$\pm$20 &  4.0$\pm$0.8 &         &        & 220 & 300$\pm$33 &  6.1$\pm$0.6 \\
1998.59 & Aug 08 &  90 & 290$\pm$10 &  7.3$\pm$0.7 &         &        & 227 & 290$\pm$14 &  5.9$\pm$0.6 \\
        &        & 150 & 240$\pm$10 &  7.2$\pm$0.7 &         &        & 230 & 280$\pm$28 &  5.7$\pm$0.6 \\
1998.60 & Aug 14 &  90 & 280$\pm$20 &  7.0$\pm$0.7 &         &        & 262 & 204$\pm$42 &  4.6$\pm$0.7 \\
        &        & 230 & 140$\pm$20 &  5.7$\pm$0.9 & 2002.96 & Apr 15 &  89 & 384$\pm$43 &  7.2$\pm$0.8 \\
        &	 &     &            &		   &         &        & 220 & 227$\pm$45 &  4.6$\pm$0.9 \\
%
2003.16 & Feb 26 &  85 & 421$\pm$22 &  8.0$\pm$0.8 &2003.21 & Mar 16 &  83 & 478$\pm$25 &  9.1$\pm$0.9 \\
        &        &  87 & 437$\pm$13 &  8.2$\pm$0.8 &        &        &  85 & 455$\pm$16 &  8.7$\pm$0.9 \\
        &        &  88 & 420$\pm$16 &  7.9$\pm$0.8 &        &        &  86 & 532$\pm$20 & 10.0$\pm$1.0 \\
        &        &  89 & 416$\pm$24 &  7.8$\pm$0.8 &        &        &  89 & 508$\pm$41 &  9.5$\pm$1.0 \\
        &        & 107 & 432$\pm$17 &  8.1$\pm$0.8 &        &        &  91 & 479$\pm$29 &  8.9$\pm$0.9 \\
        &        & 218 & 290$\pm$23 &  6.0$\pm$0.6 &        &        & 104 & 467$\pm$17 &  8.8$\pm$0.9 \\
        &        & 220 & 278$\pm$45 &  5.7$\pm$0.9 &        &        & 110 & 438$\pm$34 &  8.3$\pm$0.8 \\
        &        & 227 & 286$\pm$32 &  5.8$\pm$0.7 &        &        & 169 & 353$\pm$19 &  7.2$\pm$0.7 \\
        &        & 262 & 241$\pm$38 &  5.4$\pm$0.8 &        &        & 220 & 220$\pm$30 &  5.6$\pm$0.9 \\
        &	 &     &	    &		   &	    &        & 226 & 298$\pm$28 &  6.0$\pm$0.6 \\
        &	 &     &	    &		   &        &        & 227 & 244$\pm$30 &  6.1$\pm$0.7 \\
\noalign{\smallskip}
\hline
\end{tabular}
\end{center}
\label{SESTdata}
\end{table*}

\begin{table*}
\caption[]{JCMT observations of the Centaurus A nucleus}
\begin{center}
\begin{tabular}{lrrcclrrcc}
\hline
\noalign{\smallskip}
Day      & Date   & Frequency & Main-Beam        & Flux Density & Day    & Date   & Frequency & Main-Beam        & Flux Density \\
         &        & $\nu$     & Brightness       & S$_{\nu}$    &        &        & $\nu$     & Brightness       & S$_{\nu}$    \\  
         &        & (GHz)     &T$_{\rm mb}$ (mK) & (Jy)         &        &        & (GHz)     &T$_{\rm mb}$ (mK) & (Jy)         \\
\noalign{\smallskip}     
\hline
\noalign{\smallskip}
1997.09 & Feb 01 & 330 & 327$\pm$22 &  5.8$\pm$0.6 & 2003.36 & May 05 & 265 &            &  5.4$\pm$0.7 \\
2002.28 & Apr 12 & 230 &            &  6.3$\pm$0.9 & 2003.50 & Jul 01 & 265 &           &  4.1$\pm$0.6 \\
        &        & 268 &            &  5.9$\pm$0.9 &         &        & 330 &           &  8.4$\pm$1.3 \\
2002.47 &        & 268 &            &  6.2$\pm$0.9 & 2004.30 & Apr 17 & 345 &            &  7.2$\pm$1.1 \\
2003.24 & Mar 30 & 268 &            &  6.1$\pm$0.9 & 2004.57 & Jul 26 & 330 &           &  6.3$\pm$0.9 \\
        &        & 330 &            &  7.9$\pm$1.2 & 2005.23 & Mar 23 & 330 &           &  4.4$\pm$0.7 \\
        &        & 345 &            &  5.2$\pm$0.8 &         &        &     &           &              \\
\noalign{\smallskip}
\hline
\end{tabular}
\end{center}
\label{JCMTdata}
\end{table*}

In the time period 1989-1993, observations were made with Schottky
barrier diode receivers providing single-sideband (SSB) system
temperatures, including the sky, of 500 K (3 mm band) and 1000 K (1.3
mm band).  From 1996 onwards until the closure of the SEST in 2003,
observations were made with Superconductor-Insulator-Superconductor
(SIS) tunnel junction receivers operating simultaneously in either (i)
the 3 mm and 2 mm bands, or (ii) the 3 mm and 1.3 mm bands. Typical
system temperatures, including the sky, were 220 K in the 3 mm band
and 330 K in the 1.3 mm bands. Continuum pointings were frequently
performed on Centaurus A itself; monitoring the continuum level during
the observations provided an additional and very valuable check on
tracking accuracy.  During the long integrations, the r.m.s.  pointing
was on average on the order of 5$''$ or better. All observations were
made in double-beamswitching mode, with a 6 Hz switching frequency and
a throw of 12$'$ because this procedure yielded excellent and stable
baselines.  SEST beam sizes and main-beam efficiencies at the various
observing frequencies were listed by Israel (1992); they were
41$''$-57$''$ and 0.70-0.75 in the 3 mm band, and 23$''$ and 0.5 in
the 1.3 mm band.  Whenever feasible, we used the observatory high- and
low- resolution acousto-optical spectrometers (AOS) in parallel.

For this paper, we extracted the continuum flux density from
measurements with the low-resolution backend (bandwidth 500 MHz,
resolution 1 MHz).  To this end, we averaged, for each 5-minute
sample, the continuum levels in the 0-300 km s$^{-1}$ and 800-1100 km
s$^{-1}$ velocity intervals, straddling but avoiding the central line
emission profiles. Individual sample continuum antenna temperatures
were then converted to flux densities.  The conversion typically
ranged from 25 Jy/K at 86 GHz, 27 Jy/K at 115 GHz, 30 Jy/K at 147 GHz,
to 41 Jy/K at 230 GHz.  The flux densities listed in
Table\,\ref{SESTdata} for each observed frequency are the means of the
individual samples observed at the frequency and run listed.
Observing runs typically lasted four to eight days; the median date is
given.  In determining the means, we left out samples negatively
affected by pointing drift, which were easily recognized by a slow but
steadily decreasing continuum level.

From 1996 to 1999, the Centaurus A continuum intensity was also
monitored every few months at frequencies of 90, 150, and 230 GHz in a
dedicated program. The observations were performed in the dual beam
switch mode with a 12$'$ throw, and the data were taken simultaneously
at either 90/150 or 90/230 GHz. Individual exposures were kept short
to minimize the effect of the atmosphere and to allow a filtering of
spectra with poor quality. The r.m.s. values from the pointing runs
were typically better than 3$''$. For each of the individual spectra,
the continuum flux level was determined by fitting a zeroth-order
line.  Finally, the average was determined by a least-square fit to
the fitted continuum levels and error of all the individual
measurements.

The quality of the data is a function of the stability of the sky
(which determines the accuracy of the double-beamswitch signal
subtractions) and the accuracy of the pointing. As the pointing
depends on the proper evaluation of a series of boresight observations
taken in sequence, it is itself a function of sky stability. In
general, observations in the 3 and 2 mm windows profited from
transparent sky conditions superior to those encountered in the 1 mm
window. In addition, a particular r.m.s. pointing error has a greater
effect on smaller beamsizes, i.e. at higher frequencies.  For this
reason, observations around e.g. 90 GHz have the highest quality, and
those at 220/230 GHz are associated with larger relative errors.

\subsection{JCMT data}

From 2002 through 2005, similar molecular line observations were made
with the 15m James Clerk Maxwell Telescope (JCMT) on Mauna Kea
(Hawaii).  \footnote{The James Clerk Maxwell Telescope is operated on
  a joint basis by the United Kingdom Particle Physics and
  Astrophysics Council (PPARC), the Netherlands Organisation for
  Scientific Research (NWO), and the National Research Council of
  Canada (NRC).}. They were made in beamswitching mode with a throw of
$3'$ in azimuth using the DAS digital autocorrelator system as a
backend.  As the low declination of Centaurus A brings the source to
less than 30$^{o}$ above the horizon as seen from Hawaii, observations
were performed in timeslots covering one hour at either side of
transit.  For all spectra, we scaled the observed continuum antenna
temperature to flux densities by assuming aperture efficiencies $\eta
_{\rm ap}$ = 0.57 at 230-270 GHz and $\eta _{\rm mb}$ = 0.49 at
330-345 GHz.

\begin{figure}[]
\resizebox{10cm}{!}{\rotatebox{270}{\includegraphics*{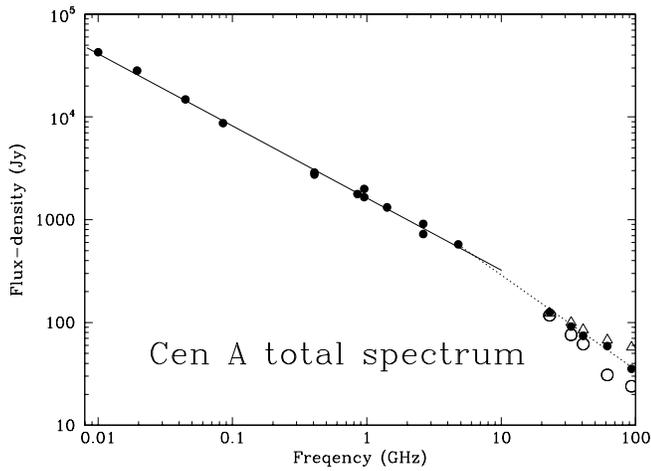}}}
\caption[] {The area-integrated continuum spectrum of the Centaurus A radio
  source from meters to millimeters. Filled circles represent
  integrated flux densities from the literature (cf. Alvarez et
  al. 2000) and from Table\,\ref{WMAPdata}. For comparison, open
  triangles represent integrated flux densities including the SE
  feature which we suspect to be unrelated Milky Way foreground (see
  Sect.~2.1) whereas open circles represent the flux densities obtained
  from Gaussian fitting of the WMAP emission which is biased to
  regions of relatively high surface brightness. Solid lines mark fits
  to the literature and WMAP data. }
\label{cena_totspec}
\end{figure}

\section{Results and analysis}

\subsection{The overall Centaurus A radio source spectrum}

In Fig.\,\ref{cena_totspec} we show the spatially integrated radio
spectrum of Centaurus A from 85 MHz to 90 GHz.  The lower-frequency
flux densities have all been taken from the literature as compiled,
corrected, and discussed by Alvarez et al. (2000), who found a
spectral index $\alpha$ = -0.70$\pm$0.01
($S_{\nu}\,\propto\,\nu^{\alpha}$) corresponding to the solid line in
Fig.\,\ref{cena_totspec}.  The higher-frequency points were taken from
Table\,\ref{WMAPdata}. A least-square fit indicates a slightly steeper
spectrum with spectral index $\alpha$ = -0.82$\pm$0.07 and a spectral
turnover by $\Delta \alpha \, \approx \, -0.12$ somewhere between 5
and 20 GHz (see also Junkes et al. 1997).  The more poorly defined
spectrum of the SW lobe is steeper than that of the NE lobe with
$\alpha_{\rm SW}\,=\,-1.0$ vs $\alpha_{\rm NE+c}\,=\,-0.6$, but this
difference is barely significant.

In these spectra and flux densities, we do not include the contribution
by the extended emission feature east of the SW lobe discussed
above. If we were to do so, its relatively strong presence at the
higher frequencies would flatten the high-frequency spectrum to the
one indicated by the open triangles in Fig.\,\ref{cena_totspec} (with
$\alpha\,=\,-0.64\,\pm\,0.09$).  The slope of that spectrum is very
similar to that at the lower frequencies, but the spectrum itself is
offset in flux density by about a factor of two and does not fit very
well onto the lower-frequency spectrum. It is clear from
Fig.\,\ref{cena_totspec} that the emission from the diffuse feature
becomes negligible compared to the Centaurus A total at frequencies
below frequencies 20 GHz. The low Galactic latitude of Centaurus A
also has some effect at much lower frequencies.  For instance, Cooper
et al. (1965) briefly discuss the presence of a relatively weak spur
of Galactic foreground synchrotron emission at the southeastern tip of
Centaurus A.  The location and extent of this spur is indicated in
their Fig.\,1.  The WMAP diffuse feature is close to, but not
coincident with, the spur and appears to be unrelated to it.  Indeed,
none of the radio continuum maps shown by Cooper et al. (1965) and
Junkes et al. (1993) in the range of 0.4 to 4.8 GHz shows emission
coincident with the WMAP feature.

There is some evidence that the ratio $R_{\rm NS}$ of the
flux densities of the NE lobes (including the central core emission)
and the SW lobes increases with frequency.  At 0.4 and 1.4 GHz, $R_{\rm
  NS}$ = 1.5 (Cooper et al. 1965, Junkes et al. 1993), but this has
increased to $R_{\rm NS}\,\approx\,3$ in the 23--41 GHz range
(Table\,\ref{WMAPdata}). Even if we were to add the suspected
foreground emission to that of the SW lobe, we still would have a
ratio $R_{\rm NS} = 2$ in the well-determined 23--43 GHz range.  We
note that Alvarez et al. (2000) have found that, over the full 0.08--43
GHz range, at least the {\it inner} lobes also have a ratio $R_{\rm
  NS}$ = 1.5. For the giant {\it outer} lobes they find a lower ratio
$R_{\rm NS}$ = 0.9, but this only applies to the limited 0.4--5.0 GHz
range.  The meaning of these apparent changes in ratio is not clear to
us.

The fraction of polarized emission is well-defined between 23 GHz and
41 GHz.  The lower-frequency, higher-resolution maps by Cooper et
al. (1965) and Junkes et al. (1993) show overall polarization 
increasing from $P/I\,\approx\,12\%$ at 960 MHz to $P/I\,\approx\,30\%$
at 5 GHz/.  We find much lower polarized fractions.  The polarization
of the entire source is within the errors constant over the 23--61 GHz
range with $P/I\,=\,6.0\pm0.3\%$.  As Fig.\,\ref{WMAPs} shows, the
{\it beam-averaged} polarization direction is different for all three
components.  The low polarization fraction found by us most likely
reflects the significant beam depolarization caused by the low resolution
of the WMAP data.

\subsection{Millimeter-wave emission of the Centaurus A core region}

\begin{figure}[]
\resizebox{9cm}{!}{\rotatebox{270}{\includegraphics*{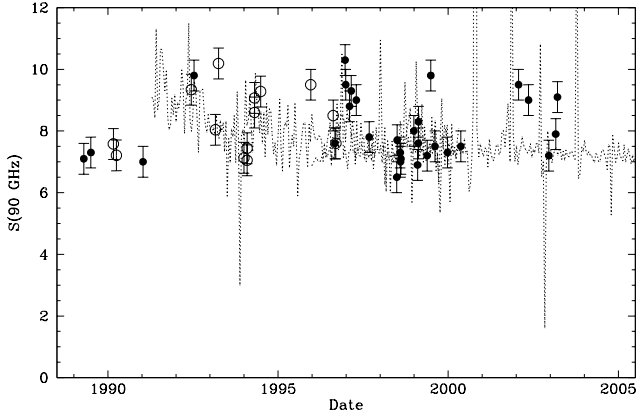}}}
\caption[] {Flux densities of the Centaurus A nucleus at 90 GHz in 1988-2005 taken 
from this paper (filled circles) and from published literature (open circles; 
see text). Dashed lines represent X-ray fluxes monitored by the BATSE and RXTE 
experiments on an arbitrary flux density scale}
\label{monitor90}
\end{figure}

\begin{table}
\caption[]{Normalized flux densities of the Centaurus A nucleus}
\begin{center}
\begin{tabular}{lcccclcccc}
\hline
\noalign{\smallskip}
Epoch   & \multicolumn{4}{c}{Flux Density (Jy)} & Spectral \\
 &  $S_{90}$ & $S_{150}$ & $S_{230}$ & $S_{345}$ & Index \\
\noalign{\smallskip}     
\hline
\noalign{\smallskip}
1989.29 &  7.1 &  --- &  --- &  --- &   --- \\
1989.50 &  7.3 &  --- &  4.5 &  --- &   --- \\
1991.03 &  7.0 &  6.6 &  --- &  --- &   --- \\
1992.53 &  9.8 &  7.3 &  5.7 &  --- & -0.58 \\
1993.55 &  --- &  --- &  7.7 &  6.6 &   --- \\
1996.67 &  7.6 &  5.4 &  --- &  --- &   --- \\
1996.97 & 10.3 &  --- & 8.2  &  --- &   --- \\
1996.99 &  9.5 &  --- & 6.65 &  --- &   --- \\
1997.10 &  8.8 &  7.5 &  --- &  5.7 & -0.32 \\
1997.15 &  9.3 &  8.7 &  --- &  --- &   --- \\
1997.30 &  9.0 &  --- &  --- &  --- &   --- \\
1997.68 &  7.8 &  6.2 &  5.1 &  --- & -0.45 \\
1998.49 &  6.5 &  9.0 &  --- &  --- &   --- \\
1998.50 &  7.7 &  6.2 &  5.1 &  --- & -0.43 \\
1998.59 &  7.3 &  7.2 &  --- &  --- &   --- \\
1998.60 &  7.0 &  6.3 &  5.7 &  --- & -0.22 \\
1998.61 &  7.1 &  6.3 &  5.7 &  --- & -0.23 \\
1998.99 &  8.0 &  6.8 &  --- &  --- &  ---  \\
1999.10 &  6.9 &  5.5 &  --- &  --- &  ---  \\
1999.12 &  7.6 &  8.5 &  --- &  --- &  ---  \\
1999.13 &  8.3 &  8.5 &  --- &  --- &  ---  \\
1999.38 &  7.2 &  6.1 &  5.3 &  --- & -0.33 \\
1999.49 &  9.8 &  8.9 &  --- &  --- &  ---  \\
1999.61 &  7.5 &  9.0 &  --- &  --- &  ---  \\
1999.98 &  7.3 &  7.5 &  --- &  --- &  ---  \\
2000.37 &  7.5 &  6.0 &  4.9 &  --- & -0.44 \\
2002.07 &  9.5 &  7.3 &  5.9 &  --- & -0.50 \\
2002.28 &  --- &  --- &  6.3 &  --- &  ---  \\
2002.36 &  9.0 &  7.0 &  5.8 &  --- & -0.50 \\
2002.47 &  --- &  --- &  6.6 &  --- &  ---  \\
2002.96 &  7.2 &  5.6 &  4.5 &  --- &  ---  \\
2003.16 &  7.9 &  6.6 &  5.7 &  --- & -0.35 \\
2003.21 &  9.1 &  7.2 &  5.9 &  --- & -0.46 \\
2003.24 &  --- &  --- &  6.5 &  6.6 &  ---  \\
2003.36 &  --- &  --- &  5.8 &  --- &  ---  \\
2004.30 &  --- &  --- &  --- &  7.2 &  ---  \\
2004.57 &  --- &  --- &  --- &  6.2 &  ---  \\
2005.23 &  --- &  --- &  --- &  4.3 &  ---  \\
\\
Means   &  8.1 &  7.1 &  5.9 & (6.1)& -0.33 \\
\noalign{\smallskip}
\hline
\end{tabular}
\end{center}
\label{standarddata}
\end{table}

\begin{figure}[]
\resizebox{11.8cm}{!}{\rotatebox{270}{\includegraphics*{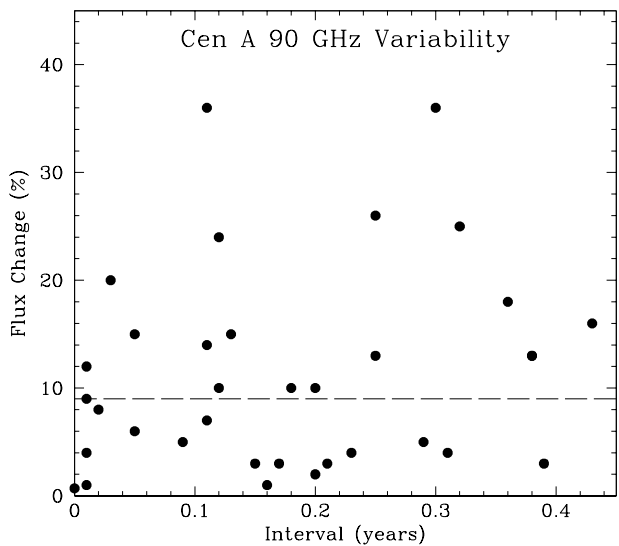}}}
\caption[] {Relative changes in 90 GHz flux densities as a function of
observing interval. Changes above the horizontal thin line are highly significant and not influenced by observing errors or bias.}
\label{cena90var}
\end{figure}

\begin{figure}[]
\resizebox{13cm}{!}{\rotatebox{270}{\includegraphics*{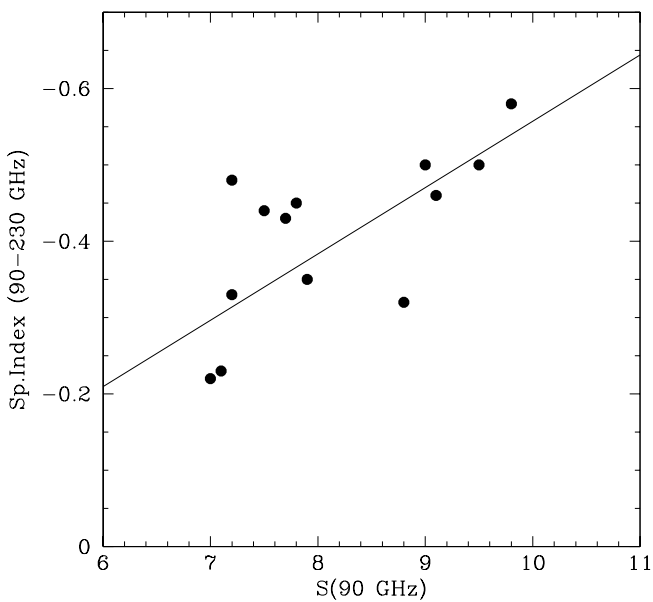}}}
\caption[] {Spectral index 90-230 GHz as a function of 90 GHz
flux density.}
\label{spectralindex}
\end{figure}

Although the extended radio emission from Centaurus A even at
millimeter wavelengths significantly exceeds that of the core region,
its surface brightness is low and declines rapidly with increasing
frequency.  Can it, however, be safely ignored in the flux density
measurements of the core with the SEST beam?  The highest frequency
measured with WMAP, 93 GHz, corresponding to a wavelength of 3.2 mm,
is in the lowest band that the SEST offered for ground based
millimeter-wave molecular line observations. The measured
gaussian-fitted flux density at this frequency is S$_{93}$(tot) =
24$\pm$5 Jy, well above the flux density $S$ = 9$\pm$1 Jy that
single-dish measurements suggest for the Centaurus A core region.
Excluding the core, the mean surface brightness at 90 GHz of the
extended synchrotron emission component is about 115 mJy/arcmin$^{2}$.
The contribution of the extended nonthermal component can thus safely
be neglected in the relatively high-resolution observations of the
core.

At very high frequencies, thermal emission from dust may also
contribute to measured intensities, but this is not the case in the
frequency range of interest to us.  The continuum emission seen at
resolutions of a few tenths of an arcminute by telescopes such as the
SEST and the JCMT, effectively originates in a point source as can
clearly be seen in the 850$\mu$m (345 GHz) SCUBA images published by
Mirabel et al.  (1999) and Leeuw et al. (2002).  From these and our
own data, we deduce that the extended emission, after correction for
20$\%$ CO line contamination, contributes less than 4$\%$ to the
central peak intensity at 345 GHz, and this contribution drops rapidly
with decreasing frequency.  In the next section we will return to this
issue.

To study the temporal behavior of the mm-wave continuum spectrum of
the core, we have reduced the measurements listed in
Table\,\ref{JCMTdata} to standard frequencies of 90, 150, and 230
GHz. We performed a linear regression on all flux density/frequency
pairs measured during a run, and for each run determined the
best-fitting `standard' flux densities by interpolation, as well as
the effective spectral index between 90 and 230 GHz. As a consequence
of this procedure, the results for each run, given in
Table\,\ref{standarddata}, are more robust than the individual
measurements taken during that run. Typical uncertainties are about
0.3 Jy at 90 GHz and 0.6 at 230 GHz. Although the sampling in time was
not uniform, we have given the mean flux density over the whole period
for each frequency at the bottom of Table\,\ref{standarddata}.  If
more detailed information is absent, these values are the best guess
for the core flux densities at any time not covered by the monitor.
We have also listed the rather flat spectral index defined by these
`best' points, with the important caveat that the three standard
flux densities defining the means were not sampled at the same times.

\subsection{Core variability at mm wavelengths}

In Fig.\,\ref{monitor90} we have plotted all observed 90 GHz flux
densities from Table\,\ref{SESTdata} and added the available (SEST)
values from the literature (Tornikoski et al., 1996; Wiklind $\&$
Combes 1997) fortunately filling in much of the gap created by our
lack of monitoring Centaurus A in the period 1993-1997.  It was clear
from the start that the core of Centaurus A is variable.  In
Fig.\,\ref{monitor90} we illustrate this variability of core flux
density.  Over the period monitored, the flux density never drops much
below 7 Jy, but at times may be higher than 10 Jy.  Sustained periods
of relatively high flux densities appear to have occurred in
1993/1994, 1997, and 2001/2002.  There is no doubt that this
variability is real. The excursions exceed the uncertainty associated
with individual plotted values, and the behavior is similar at all
observed frequencies.  At 90 GHz, the core flux density is seen to
change by as much as 20$\%$ in a few weeks.  This is illustrated in
Fig.\,\ref{cena90var}, which plots the percentage change from one
observing date to another as a function of the elapsed time.  Over the
observed period, the greatest changes (more than 25$\%$) take place
over time intervals typically between 2 and 4 months.  The
high-frequency variability observed here should be compared to the
variability deduced from the lower frequency monitoring observations
by Botti $\&$ Abraham (1993) and Abraham (1996).  Although there is,
unfortunately, only limited temporal overlap, the pattern is very
similar, with significant and sometimes rapid intensity changes. In
particular, enhanced radio emission is seen in 1992 in both our sample
and in the data by Abraham (1996). The 43 GHz variability has a large
amplitude of 8 -- 10 Jy (corresponding to factors of two change in
brightness) and is about half as much (factor of 1.25) at 22 GHz.  We
note that those flux densities were determined with relatively large
beams of 2.1$'$ and 4.2$'$, allowing significant contributions to the
emission especially at 22 GHz, not related to the core.

The X-ray variability of the Centaurus A nucleus was monitored over
the same period in the 20--200 keV band by the BATSE experiment
onboard the Compton Gamma Ray Observatory from 1991 through 2000 and
in the 2--10 keV band by the RXTE All Sky Monitor from 1996 onwards.
We took the public data from the NASA Goddard SFC and MIT websites and
constructed averages over 15-day bins. The results are plotted with
the 90 GHz flux densities in Fig.\,\ref{monitor90}.  The
correspondence between the SEST 90 GHz and the BATSE X-ray flux
densities is quite good, suggesting closely related, perhaps even
identical, radiative processes .  The correlation between the
millimeter-wave continuum and the relatively soft X ray emission
sampled by the RXTE experiment is much poorer.  Minor X-ray excursions
(on the order of a factor of two) occurred in 1991/1992, during 1994,
in late 1996 and early 1997, during 1999, and 2001. They correlate
with radio maxima, although unfortunately we are lacking millimeter
continuum coverage for 2001.

Although the behavior is the same at all frequencies observed by us,
flux densities vary less at higher frequencies.  As a result, the core
spectral index changes with the degree of activity.  In
Table\,\ref{standarddata} we have calculated spectral index values
$\alpha$ ($S_{\nu}\propto\nu^{\alpha}$) for the frequency range of 90
GHz to 230 GHz. To guarantee a reasonable accuracy (on the order of
0.10 or better in the index), we included only epochs at which
measurements were available in three different frequency windows. In
Fig.\,\ref{spectralindex} we have plotted these spectral indexes as a
function of the 90 GHz flux density.  It is quite obvious that the
mm-wave spectrum of the Centaurus A core becomes steeper (smaller
spectral index) when the actual flux increases; i.e. increases are
less in both absolute and relative senses at the higher
frequencies. The points in Fig.\,\ref{spectralindex}, with their
uncertainties of $\Delta\alpha\approx0.10$, can be fitted with a
linear-regression line $\alpha$ = -0.087 $S_{90}$ + 0.311. For the
observed lower limit of 7 Jy to the Centaurus A core flux density,
this fit implies a fairly flat spectral index $\alpha$ = -0.3
(although the two actually measured values imply an even flatter
spectrum with $\alpha$ = -0.22). Spectral indices $\alpha$ = -0.6 are
approached at enhanced flux densities of 10 Jy.

Variations in the spectral index between 22 and 43 GHz were likewise
deduced by Botti $\&$ Abraham (1993) and plotted as a function of time
in their Figure 2. If we plot their data again, now with the spectral
index as a function of the relatively strongly varying 43 GHz
flux density, we find a behavior {\it opposite} to that deduced by us
for the 90 -- 230 GHz range.  With relatively little scatter,
$\alpha_{22-43}$ = + 0.11 $S_{43}$ - 2.2; i.e.  the spectrum {\it
  flattens} when $S_{43}$ increases!  We do not have a full
explanation for this, but some remarks are in order. First, we note
that the 43 GHz flux density appears to vary even more than the 90 GHz
flux density.  Although there is very little temporal overlap, this is
nevertheless consistent with our conclusion that the spectrum steepens
when flux densities increase.  Second, the 22 GHz emission would be
expected to vary even more, but obviously fails to do so.  This could
be explained by high optical depths of the variable emission component
at this frequency.

\subsection{Nature of the core emission}

In the quiet state, the core of Centaurus A has a high-frequency spectral
index $\alpha_{high}$ = -0.3, but it must become optically thick
around 80 GHz to allow flux densities to be a factor of two
lower at 43 GHz than at 90 GHz.  This implies that the emission at 22
GHz must arise almost entirely from non-core components. In the period
covered by our observations, Centaurus A has not shown much activity, in any
case less than in earlier decades (cf Abraham$\&$ Botti
1993). Nevertheless, we did observe a few episodes with slightly
enhanced activity. In these mildly `active' states, flux densities
increase, the high-frequency spectral index steepens to
$\alpha_{high}$ = -0.6, and the spectral turnover shifts to lower
frequencies of about 45-50 GHz.  In this state, there is a measurable
2--4 Jy contribution from the core to the total 22 GHz emission from
the central 4 arcmin, such as the 4 Jy seen in the VLBI measurements
by Tingay et al. during the 1995 episode, but most of the 22 GHz
emission still represents non-core components.  

Recent (April 2006) observations with the Smithsonian Millimeter Array
(SMA) at a resolution of $2''\times6''$ show a point source with a
continuum flux density $S_{230GHz}$ = 6 Jy (Espada et
al. 2007).  Although we do not have a single-dish measurement at the
same epoch, a glance at Table\,\ref{standarddata} will show that this
is, in effect, the total flux to be expected at this frequency.

What does the observed core emission represent, and what is the
meaning of the variability?  Although our observing beams were much
larger, essentially all observed emission arises from a region with a
size of a few arcsec at most, as indicated by the SMA observations
just mentioned, as well as the measurements described below.  VLBI
images at frequencies of 5 and 8 GHz (Tingay et al. 1998, 2001; Tingay
$\&$ Murphy 2001; Horiuchi et al. 2006) show very considerable source
structure concentrated within about 60 milliarcsec.  We note that the
{\it submillimeter core itself} suffers from both free-free absorption
and synchrotron selfabsorption at these frequencies (see e.g. Tingay
$\&$ Murphy 2001).  Nevertheless, the images suggest that the emission
observed by us arises from a nuclear source, as well as from bright
nuclear jets extending over the better part of a parsec.  This
complicates the interpretation of the observed variability, since we
have no means of determining where precisely it originates.  The time
sequences registered by Tingay et al. (1998, 2001) strongly suggest
that nuclear jet evolution is a major source of variability in the
emission, as total 8 GHz flux densities vary between 4 and 10 Jy,
Their maps also show movement and variability of individual components
in the nuclear jets.  The 8 GHz `core' component varies much less
(between 1.8 and 3.3 Jy with a median of 2.4 Jy over the period
1991-2000) and the 8 GHz and 22 GHz core flux densities are almost
identical at the only two epochs (1995.88 and 1997.23) for which a
reliable measurement exists (Tingay et al. 1998, 2001).  However, in
the much higher resolution (1.2$\times$0.6 mas) maps by Horiuchi et
al. (2006) this `core' breaks up into at least four separate
components.  These extend over 7 milliarcsec (0.13 pc) and should
presumably be considered as inner nuclear jet components. It is thus
unclear whether the actual nuclear source (if any) itself is variable
or not.

Abraham et al. (2007) have interpreted rapid, day-to-day variations
with amplitudes of 20$\%$ in the 43 GHz emission observed over a
three-month period in 2003 as evidence of free-free absorption of the
nuclear source by clouds in the center of NGC~5128.  It is, however,
by no means clear that such an explanation could also apply to
longer term variations, or indeed to the pattern of variability at
higher frequencies.

Figure\,\ref{spectralindex} shows that the core spectrum steepens when
it brightens, implying rapid energy loss of any electrons injected
during active periods.  We might assume that the emission from the Cen
A core consists of a variable component and a component of constant
emission.  In that case, the spectral index of the variable emission
is reasonably well-constrained with $\alpha_{var}$ = -1.6, especially
if the 43 GHz peak excursions of about 9 Jy (Abraham 1996) correspond
to the 3 Jy peak excursions at 90 GHz, which we do not know for certain.
The relative flux densities of the two components and the spectral
index of the constant component need an additional constraining
assumption.  The lowest flux densities measured at 43 GHz are 3-4 Jy,
which puts an upper limit on the constant component.  For instance, if
we assume a flat spectrum for the latter ($\alpha_{cst}$ = 0.0), the
minimum flux density of the variable component is $S_{90}^{min}(var)$
= 3 Jy.  However, there does not seem to be a consistent model for the
variation at all three frequencies without additional and ad-hoc
assumptions.  It appears that millimeter-wave VLBI monitoring is
needed to provide the definitive answer to the question of the
Centaurus A nuclear source properties.  In particular, it would be
quite interesting to compare the results of such monitoring with the
results for the core of Centaurus A extracted by Meisenheimer et al.
(2007) from mid-infrared interferometry at the ESO-VLT.  After
completion, the maximum resolution of the Atacama Large Millimeter
Array (ALMA) will be about 50 mas at 100 GHz. This will close the gap
between the centimeter-wavelength synthesis array {\it resolutions}
and the centimeter-wavelength VLBI {\it fields of view} noted by
Israel (1998, Sect. 5.4), and thus be invaluable for studies of the
evolution of the nuclear and inner jets.  The resolution falls far
short, however, of those needed to separate processes occurring in the
nuclear jets and in the core components.  If THz imaging is feasible
at the longest baselines, the resulting ALMA resolution of 5 mas is
better suited to such studies.  ALMA sensitivities will be amply
sufficient to show the core, but it remains to be seen how much of the
subparsec jet structure will be discernible.

\section{Conclusions}

We have extended the spatially integrated continuum spectrum of the
Centaurus A radio source by using WMAP results between 23 and 93
GHz. The spectrum has now been explored over four frequency decades,
from 10 MHz to 93 GHz.  Between 10 MHz and 5 GHz, the spectral index
$\alpha\,=\,-0.70$, as determined in the literature.  Longwards of 5
GHz the spectrum appears to steepen, with
$\alpha\,=\,-0.83\,\pm\,0.07$.  The flux density ratio $R_{\rm NS}$ of
the NE lobes (including the core) and the SW lobes increases with
frequency from $R_{\rm NS}\,=\,1.5$ at 0.4--1.4 GHz to
$R_{NS}\,\approx\,3$ in the 23--61 GHz range.

We monitored the emission from the Centaurus A core component at
frequencies between 90 and 230 GHz over more than ten years.  The core
is variable in emission, but over the period 1989--2003, the 90 GHz
flux density from the compact core was never below 7 Jy. The spectrum
of the quiescent core is relatively flat with $\alpha$ = -0.3.
Variability was less at higher frequencies, implying steepening of the
continuum spectrum simultaneous with core brightening.  During active
periods, the emitted flux increases, but the spectral turnover
frequency decreases. The nuclear component is optically thick below
45--80 GHz.

It appears that most if not all of the variability is associated with
the inner nuclear jet components that have been detected in VLBI
measurements, but the mechanism of variability is not yet clear.

\acknowledgements{It is a pleasure to thank C.L. Bennett for
  communicating Centaurus A WMAP fluxes at an early stage, N. Odegard
  for insightful advice on the extraction of fluxes from WMAP data
  products, M.P.H. Israel for valuable assistance in data handling, an
  anonymous referee for comments leading to improvements in the paper,
  and H. Steinle's website
  (http://www.gamma.mpe-garching.mpg.de/~hcs/Cen-A/) for its very
  useful information.}


\begin{thebibliography}{}
%
\bibitem{} Abraham Z., 1996 in: Extragalactic Radio Sources, IAU Symposium 175, R. Ekers et als (eds), p. 25
\bibitem{} Abraham Z., Barres de Almeida U., Dominici T.P., $\&$ Caproni A., 2007 \mnras 375, 171
\bibitem{} Alvarez H., Aparici J., May J., \& Reich P., 2000 \aua 355, 863
\bibitem{} Bennett C. L., Bay M., Halpern M., and 12 coauthors, 2003a, \apj 583, 1
\bibitem{} Bennett C. L., Halpern M., Hinshaw G., and 18 coauthors, 2003b, \apjs 148, 1
\bibitem{} Bennett C. L., Hill R. S., Hinshaw G., and 13 coauthors, 2003c \apjs 148, 97
\bibitem{} Bolton J.G., $\&$ Clark B.G., 1960 \pasp 72, 29
\bibitem{} Botti L.C.L., $\&$ Abraham Z., 1993 \mnras 264, 807
\bibitem{} Burns J.O., Feigelson E.D., $\&$ Schreier E.J., 1983\apj 273, 128
\bibitem{} Cooper B.F.C., Price R.M, $\&$ Cole D.J., 1965 Aust. J. Phys. 18, 589
\bibitem{} Combi J.A., $\&$ Romero G.E., 1997 \auas 121, 11
\bibitem{} de Mello D.F., $\&$ Abraham Z., 1990 Rev. Mex. Astr. Ap. 21, 155
\bibitem{} Fogarty W.G., $\&$ Schuch N.J., 1975 Nature, 254, 124
\bibitem{} Fujisawa K., Inoue M., Kobayashi H., and 10 coauthors, 2000 \pasj 52, 1021
\bibitem{} Hawarden T.G., Sanell G., Matthews H.E., et al. 1993 \mnras 260, 844
\bibitem{} Horiuchi S., Meier D.L., Preston R.A., $\&$ Tingay S.J., 2006, \pasj 58, 211
\bibitem{} Israel F.P., 1998 \aar 8, 237
\bibitem{} Hinshaw G., Weiland J.L., Hill R.S., and 18 coauthors, 2008, \apjs in press, (arXiv:0803.0732)
\bibitem{} Junkes N, Haynes R.F., Harnett J.I., $\&$ Jauncey D.L,  1993 \aua 269, 29 (erratum 1993 \aua 274, 1009)
\bibitem{} Junkes N., Haynes R. F., $\&$ Mack K.-H., 1997 Astron. Ges., Abstr. Ser., No. 13, p. 67
\bibitem{} Kellerman K.I., Zensus J.A., $\&$ Cohen M.H., 1997, \apjl 475, L93
\bibitem{} Leeuw L.L., Hawarden H.G., Matthews H.E., Robson E.I., $\&$ Eckart A., 2002 \apj 565, 131
\bibitem{} Meier D.L., Jauncey D.L., Prestion R.A., and 20 coauthors, 1989 \aj 98, 27
\bibitem{} Meisenheimer K., Tristram K.R.W., Jaffe W., and 11 coauthors, 2007 \aua 471, 453
\bibitem{} Mirabel I.F., Laurent O., Sauvage M., 1999 \aua 341, 667
\bibitem{} Sheridan K.V., 1958 Austral. J. Phys. 11, 400
\bibitem{} Tateyama C.E., $\&$ Strauss F.M., 1992 \mnras 256, 8 
\bibitem{} Tingay S.J., Jauncey D.L., Reynolds, J.E., and 23 coauthors, 1998 \aj 115, 960
\bibitem{} Tingay S.J., $\&$ Murphy D.W., 2001 \apj 546, 210
\bibitem{} Tingay S.J., Preston R.A., $\&$ Jauncey D.L., 2001 \aj 122, 1697
\bibitem{} Tornikoski M., Valtaoja E., Ter\"asranta H., et al, 1996 \auas 116, 157
\bibitem{} Wiklind T., $\&$ Combes F., 1997 \aua 324, 51
\bibitem{} Wright E.L., Chen X., Odegard N, and 18 coauthors, 2008, \apjs, submitted (arXiv:0803.0577)
\end{thebibliography}
\end{document}